# Poloidal Inhomogeneity of the Particle Fluctuation Induced Fluxes near of the LCFS at Lower Hybrid Heating and Improved Confinement Transition at the FT - 2 Tokamak.

S.I. Lashkul, S.V.Shatalin*, A.B. Altukhov, A.D. Gurchenko, E.Z. Gusakov, V.V. Dyachenko, L.A.Esipov, M.Yu. Kantor, D.V. Kouprienko, A.Yu. Stepanov, A.P. Sharpeonok, E.O.Vekshina*.

*A.F.Ioffe Physico-Technical Institute, Politekhnicheskaya 26, 194021, St.Petersburg, Russia, Serguey.Lashkul@pop.ioffe.rssi.ru,*
**St.Petersburg State Polytekhnical University, St.Petersburg, Russia,*

## Introduction

The problem of the anomalous high periphery thermal transport at tokamak plasmas and mechanism of their suppression, when a transport barrier is formed has been discussed in a number of recent publications [1, 2]. This paper present our observations and conclusions about development of the transport process at the plasma periphery of the small tokamak FT-2 during additional Lower Hybrid Heating (LHH), when external (ETB) transport barrier followed by Internal (ITB) transport barrier is observed [3, 4].

The main part of the previous papers is concerned with plasma under $q = 6$ (R=0.55m, a=0.079m, $I_{pl} = 22$kA and $B_t = 2.2$T, $P_{LHH} = 90 \div 100$kW), where the effective LHH and improved confinement transition are realised [3]. The L-H transition with ETB has been observed after RF pulse end. The RF pulse ($\Delta t_{LH} = 5$ms) is applied at the 30$^{th}$ ms of a $\Delta t_{pl} = 50$ms plasma shot. During additional LHH the ITB is formed spontaneously a few ms after the RF pulse start. The essential gradient rise of the ion temperature profiles and its flattening in the core center during LHH indicates that ITB is formed at 32 – 33 ms in the r = 5 cm region. The transport barrier formation can be confirmed by time history of the ion temperature variation obtained by CX analyser and spectral diagnostic using Doppler broadening of the CIII line radiation and density profile changes [4].

It should be stressed that in our papers effective LH heating and spontaneous ITB formation have been explained through careful numerical Monte Carlo modelling [5]. One deduced that the high energy ions are well confined in the case, when poloidal Mach number $M_p = \ \geq 1$, which has been realised in FT-2 experiment with plasma under $q = 6$. It was found the radial electric field provides a spontaneous transition to high negative value, if the local Mach number is about one. This is partially due to rotational runaway due to negative inertia [6], and partially due to finite orbit effects [7].

The L-H transition with ETB has been observed after RF pulse end. Distinction of the mechanism which can trigger L – H transition is discussed in this paper. The paper deals with the new microturbulence and spectral experimental data and their analysis, which show, that the radial electric field $E_r$ generated at the LH heating (LHH) in the FT-2 is high enough to form the transport barriers.

## Experiment

I.

The paper deals with the structural feature of the Particle Fluctuation Induced Fluxes near of the Last Close Flux Surface (LCFS). The peculiarities of the variations of the fluctuation fluxes near periphery are measured by three moveable multielectrode Langmuir probes (L-probe) located in the same poloidal cross-section of the chamber. The movable multielectrode L-probes allowed us to study the behaviour of the local electron temperature, plasma density, spatial potential, electric field, as well as 2D (poloidal cross-section) $E\times B$ drift flux densities with respect to time practically at any poloidal angle [8]. The measurements were carried out shot by shot with a radial (1 mm step size) and poloidal angle spacing. The $E\times B$ drift flux can be expressed by the sum of the quasi-steady-state $\boldsymbol{\Gamma}_0(t)=cn_0(t)[\mathbf{E}_0(t),\mathbf{B}]/B^2$ and fluctuation-induced $\boldsymbol{\Gamma}^{\sim}(t)=c[\langle n^{\sim}(t)\mathbf{E}^{\sim}(t)\rangle,\mathbf{B}]/B^2$. The radial fluctuation-induced flux component can be expressed as $\Gamma_r^{\sim}(t) = C_{n(\sim)E(\sim)}c\langle n^{(\sim)2}(t)\rangle^{1/2} \langle E_\theta^{(\sim)2}(t)\rangle^{1/2}/B_\varphi$, (angular brackets indicate averaging with respect to time). Here $C_{n(\sim)E(\sim)}$ is the correlation coefficient [1, 11]. To obtain the quasi-steady-state $\boldsymbol{\Gamma}_0(t)$ flux the low frequency (smoothing) probe signal component was used. The turbulent $\boldsymbol{\Gamma}^{\sim}(t)$ flux is determined by the digital or analogous treatment of the probe signal in the wide frequency (10 - 500 kHz) range. Experiment shows that in the SOL plasma parameters depend not only on the minor radius, but also on the poloidal angle θ. For example, the maximum values of the electron temperature (up to 40 eV at the LCFS) were observed at angles of $\theta = 90^0 \ldots 150^0$ and $220^0 \ldots 250^0$ where the density is maximal. During LHH, the temperature distribution became more flattened in the poloidal direction. In the post-heating phase (H-mode), the distribution of $T_e$ (up to 20 eV) was strongly inhomogeneous with a pronounced maximum at Low Field Site (LFS), $\theta = \pm 10^0$. Similar distributions of $T_e$ were observed also in the deeper limiter shadow regions, which indicates that the plasma parameters in the SOL are governed not only by radial heat fluxes, but also by other factors (e.g., poloidal particle fluxes) [8]. One also notes that the radial profile of $T_e$ at same poloidal angels can be non-monotonic. One should draw attention, that complicated energy balance of the SOL depends on toroidal drift and longitudinal thermo-conductivity along magnetic lines. Additionally one should take into account the small shift of the plasma core (a few mm) along of the major radius during the LHH [8, 9].

II.

To measure accurately the change of main periphery parameters during one shot we used specific L-probes localisation. One probe was located at the low field side with its axis inclination in respect to the equatorial plane $\theta = 60^0$, and the second one was located at the high field side with inclination $\theta = 230^0$ [10]. The reading of the θ value is taken from the outward equatorial plane into the electron diamagnetic drift direction (the upward part of the cross-section). The shape of the head of the probes and the inclination of the probes permits to measure directly the vectors of the poloidal and radial electrical fields at once. In such a case the poloidal and radial components of the electrical field fluctuations as well as the density fluctuations (saturated probe current), have been registered. This data permits direct measurements of the radial and poloidal fluctuation fluxes during a single discharge with subsequent calculation of the quasi-steady-state fluxes.

The first results were obtained, that permitted to make the following conclusions [10]: (i) the data demonstrate a number of both common and distinct features presenting the poloidal inhomogeneity of the poloidal electric field and density fluctuations at the inner and outer perimeter of the toroidal plasma core. (ii) The fluctuation particle fluxes consist of discrete blobs, which move as filaments. (iii) The blobs at high field side $\Theta=230^0$, r =7.4 cm is directed mainly outward. The frequency of the spikes and their dimensions in the post-heating phase (in H-mode) are lower essentially than in OH (L-mode) period. The fluctuation fluxes at low field side $\Theta = 60^0$ are more non-uniform. That is confirmed by the change of flux direction from outward to inward during RF pulse at r =7.9 cm. (iv) The measured shear of the radial electric field was discussed as the mechanisms resulting in suppression of the anomalous particle transport. (v) The reduction in turbulence induced radial particle transport is due to both de-correlation of the $E^{(\sim)}{}_{\boldsymbol{\theta}}$ and $n^{(\sim)}$ and changes in the amplitudes of $E^{(\sim)}{}_{\boldsymbol{\theta}}$. (vi) The absolute $E^{(\sim)}{}_{\boldsymbol{\theta}}$ fluctuation level shows dependence on the sign of $E_r$ shear. The sharp suppression of $E_\vartheta$ fluctuation is registered when $E_r$ becomes negative and shear of $E_r$ has the positive value, which is observed immediately after the end of RF pulse. A sharp decrease of the density fluctuation is supported by the X-mode reflectometry also after RF pulse at $r = 7.5cm$ [4]. In this experiments we noted that fluctuation-induced $\boldsymbol{\Gamma}^{\sim}(t)$ and the quasi-steady-state $\boldsymbol{\Gamma}_0(t)$ drift fluxes make nearly equal contributions to the radial transport, while the input of the fluctuations in the poloidal plasma flow is negligible in comparison with $\Gamma_0$ [6].

III.

So, the abrupt rise of the negative $E_r$ at the end of RF pulse could result in L-H transition. To reveal the mechanism of this effect we have recently repeated the experiment (#260504) at the same scenario with LHH start at 30$^{th}$ ms. One L-probe was located at the low field side with its axis inclination in respect to the equatorial plane $\theta = 310^0$, and the second one is located at the high field side with inclination $\theta = 230^0$. The plasma core was slightly shifted upward (at $a \cong 77$ mm) for measurements by probes in the deeper layers and for some adjustment of the poloidal parameters symmetry. The scenarios with LHH are the same as mentioned above [10], when ITB and ETB are formed. The Figures 1 and 2 demonstrate changes of the $T_e(r, t)$ and $n_e(r, t)$ profiles during LHH near the LCFS for two poloidal angles. The L-probes are inserted deeper shot by shot into the plasma core up to 74 mm. The SOL and limiter shadow region run from ~79 mm to 81 mm and onward. For two poloidal angles one can see, that $T_e$(r=74÷76mm) decrease faster after 33 ms in comparison with $T_e$ in SOL, where longitudinal conductivity and poloidal fluxes could be dominant. The $E_r= -\delta\varphi_f/\delta r-(3/e)(\partial T_e/\partial r)$ and quasi stationer poloidal flux $\Gamma_{0\theta}$ are shown in Figures 3 and 4. The Figures illustrate fast rise of negative value of the $E_r$ and therefore $\Gamma_0$ after the end of RF pulse for two poloidal L-probe position. One can see the appearance of the large poloidal quasi-steady-state flux shear and the point of reverse directions of the fluxes, which can stimulate ETB formation with large density gradient observed after the end of RF pulse at 35 ms. The time history changes of the plasma periphery parameters $N_{L,e}(74mm)$, $E_\theta\tilde{}$, correlation coefficient $C_{n(\sim)E(\sim)}$ (for $r$ = *74-77mm*), are shown in the Fig. 5 for L-probe ($\theta = 230^0$ , HFS). $H_\beta$ radiation and small shift plasma column $\Delta R$ are illustrated also. This data show, that during of the LHH (at 32 – 33ms), when ITB is formed on $r$ = *4-5cm*, $C_{n(\sim)E(\sim)}$ gradually decreases at all marked radii, the same as $H_\beta$ radiation decreases. De-correlation effect runs from plasma core and obviously leads to the decrease of the $E_\vartheta\tilde{}$, $N\tilde{}$, therefore the directed outward particle fluctuation-induced flux $\boldsymbol{\Gamma}\tilde{}(t)$ decreases. For L-probe at Low Field Fide (LFS), ($\theta = 310^0$) the same data demonstrate tendency for $\boldsymbol{\Gamma}_r\tilde{}(t)$ to change their outward direction to inward, that is obviously associated with the RF launched wave effect (Fig. 6).

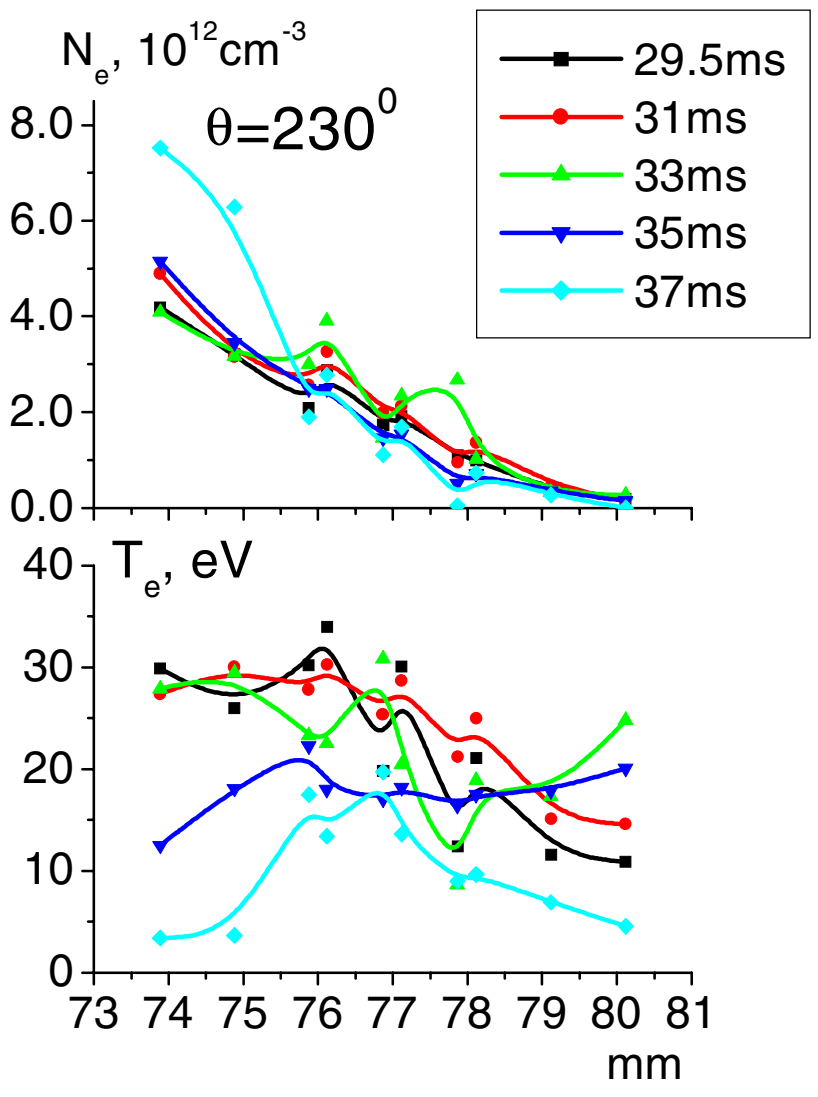

*Fig. 1*

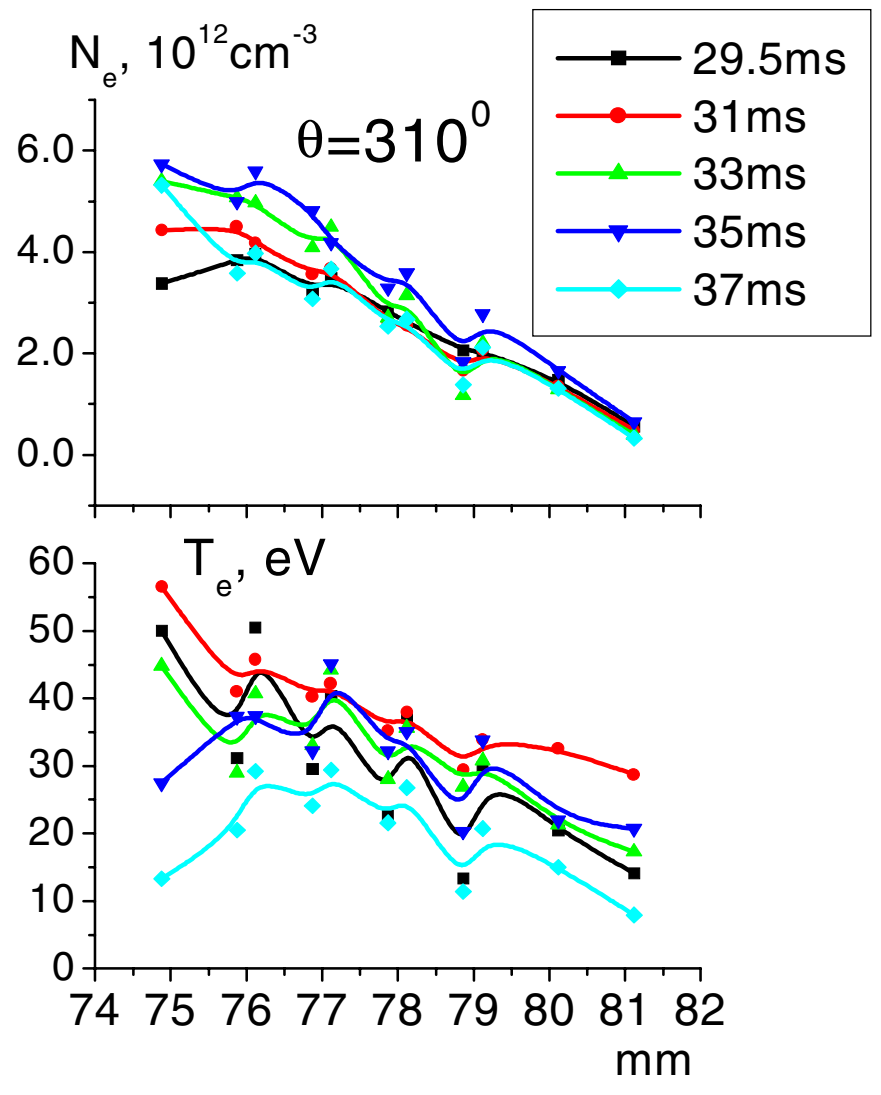

*Fig. 2*

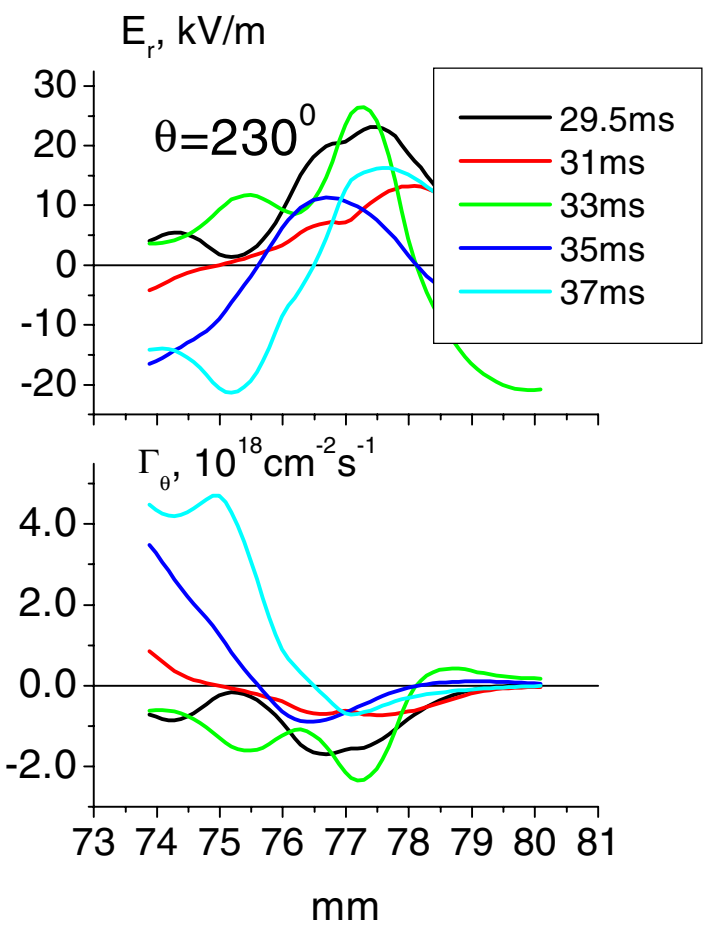

*Fig. 3*

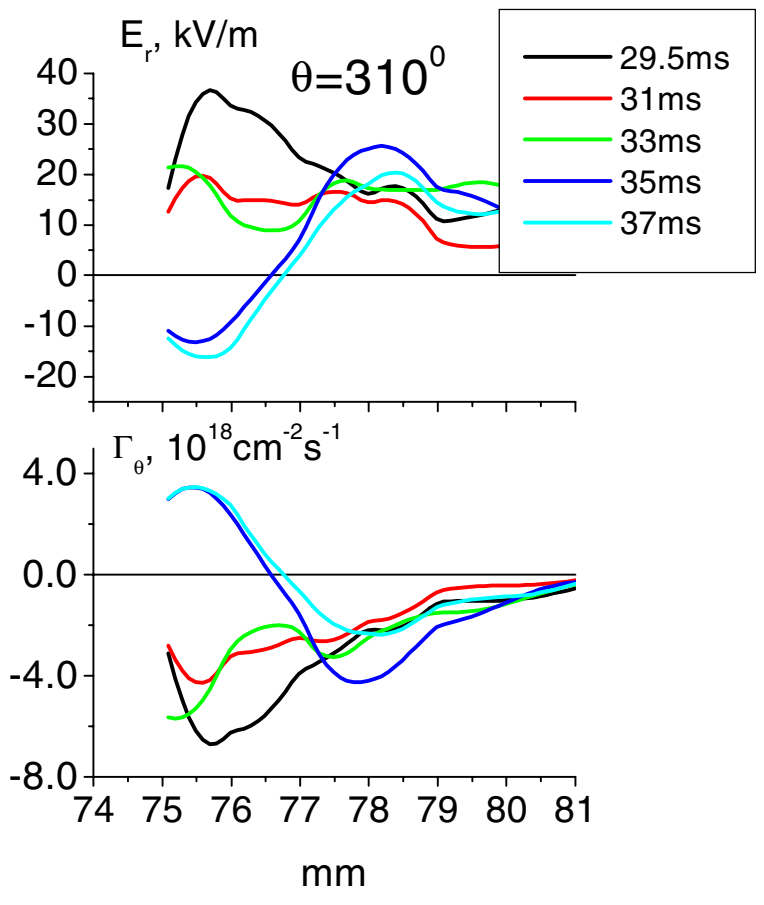

*Fig. 4*

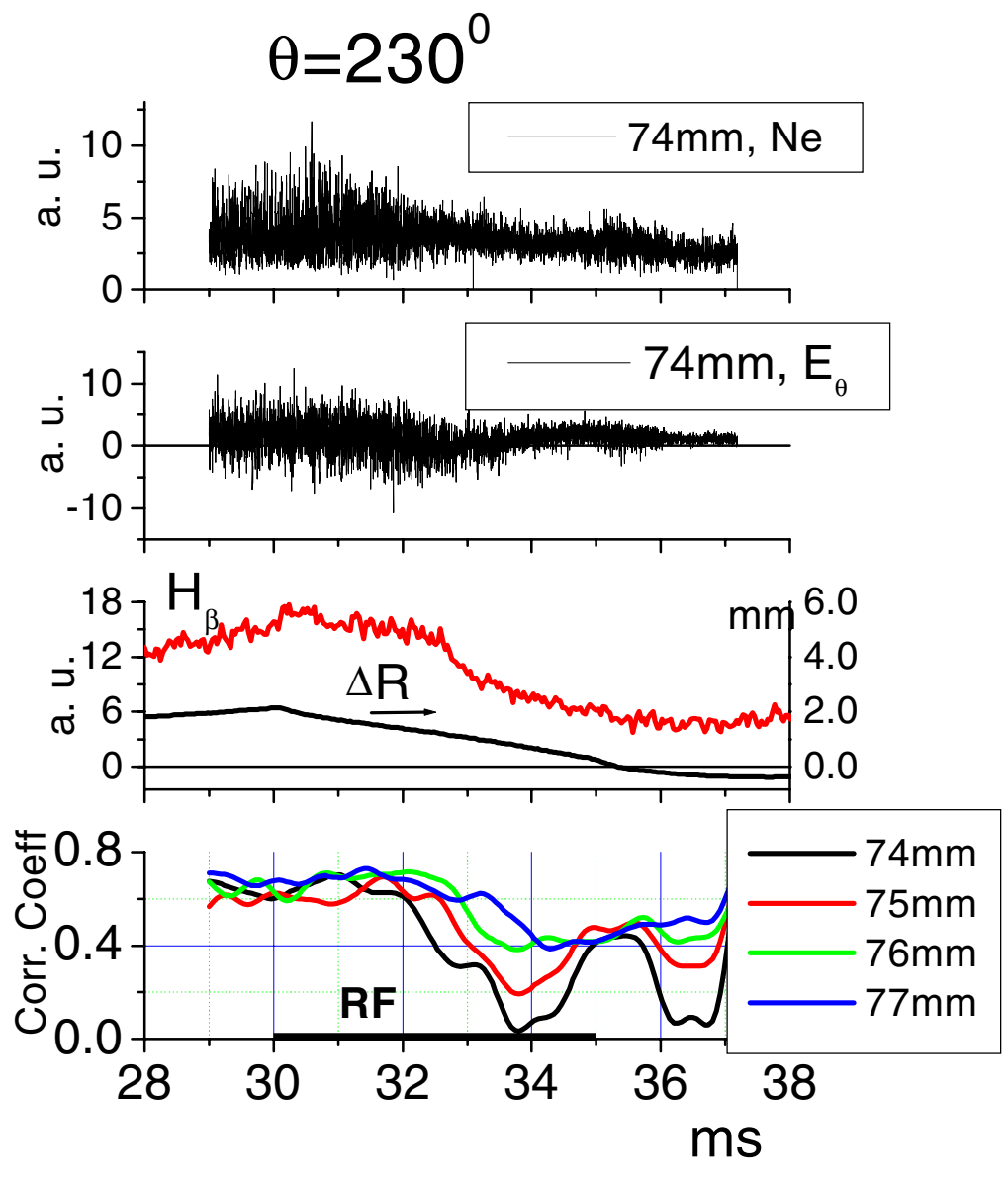

*Fig. 5*

The X-mode reflectometry measurements (#020604) have been repeated with a new steerable double antennae set (designed for UHR backscattering (BS) [12]) vertically shifted by 15 mm from equatorial plane. New quadrature detection scheme provides Doppler frequency ($f_D$) shifted spectra measurements of density fluctuations. The simple 2D ray tracing code was used for poloidal wave number ($k_\vartheta$) calculations in the BS point near cut off layer for poloidal rotation velocity estimations: $V_\vartheta = \pi f_D / k_\vartheta$. First measurements did not show any decrease of the turbulence level at the gradient region $r = 5 \div 6.5$ cm during LHH, but they pointed out the significant rise of $V_\vartheta(r = 5\text{-}6$ cm) from ~ 0 up to ~3 km/s during RF pulse in the electron diamagnetic drift direction, in a good agreement with $V_{\vartheta,E\times B}$ data obtained by spectral measurements using CIII line. Observed $V_\vartheta$ rise could perhaps trigger the ITB formation at 32–33 ms. Due to the sizeable antenna diagram width in the near zone the interpretation of the reflectometry data at the plasma edge ($r = 6.5$ - 8 cm) is not trivial and needs in future the implementation of the full wave modeling for taking into account the interference between probing and scattered radiation.

**Conclusion**

So the observed L – H transition and ETB formation after LHH and the associated negative $E_r$ rise [10] are resulted mainly from the decrease of the $T_e$ near inner region of the LCFS by greater extent than in SOL. This effect is stimulated by decrease of the input power and decrease of radial $C_{n(\sim)E(\sim)}$ (for $r = 74\text{-}77mm$) (and radial particle fluctuation-induced $\Gamma^{\sim}(t)$) resulted from ITB formation mechanism during LHH [4]. $T_e$ variation in the SOL after LH heating pulse takes place to a lesser extent. Observed nonmonotonic radial profile of $T_e$ near LCFS with positive $\partial T_e/\partial r$ *rise* is kept constant obviously by large longitudinal conductivity and poloidal fluxes from the hotter limiter shadow regions because of the poloidal inhomogeneity of the $T_{e\ SOL}$ and $n_{e\ SOL}$ [8]. Such induced negative $E_r$ after RF pulse gives fast rise to a quasi-steady-state $\Gamma_0(t)$ drift fluxes with reversed direction structure, like “zonal flows” [13], which may inhibit transport across the flow. Large rise of grad($n_e$) after LHH near LCFS with L-H transition is observed after the end of LH pulse for a long time – about 10-15ms.

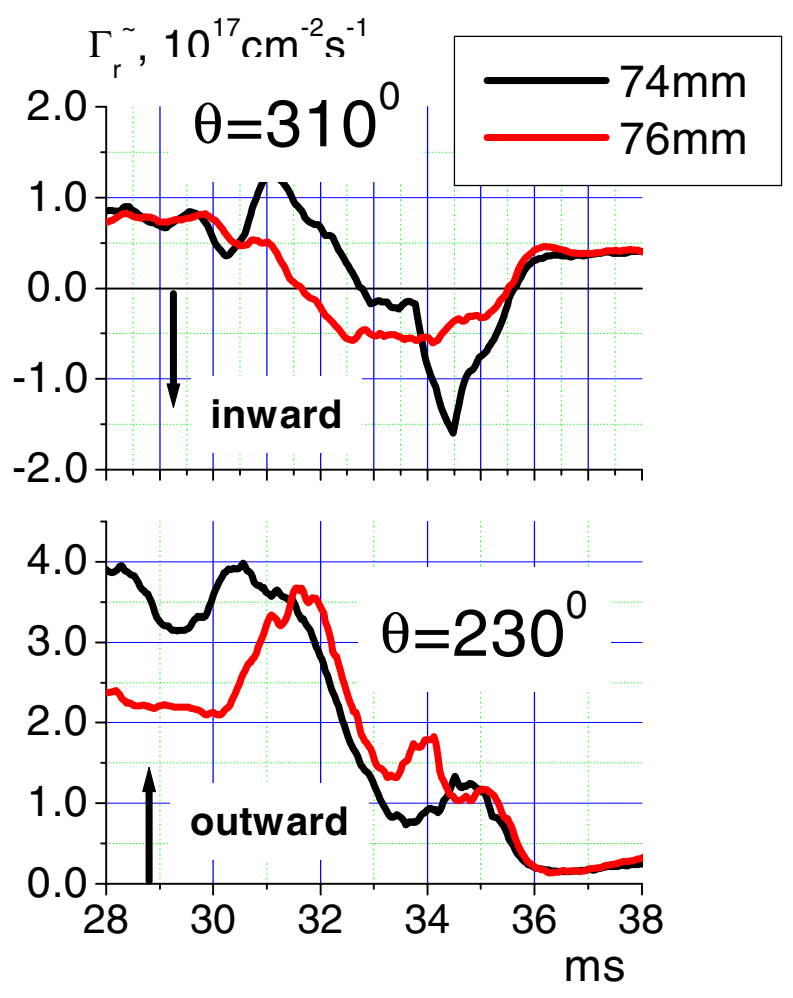

*Fig. 6*

The study was performed with the support of INTAS-01-2056, YSF2002-104 and the Russian Basic Research Foundation Grants 02-02-17684, 02-02-17591, 02-02-17589, 04-02-16534; TO 2-7.4-2694 grant and RF Leading Scientific Schools grant 2216.2003.2.

***References***

1.J. Boedo et al. Nuclear Fusion, Vol. 40, No. 7 (2000) 1397 - 1410

2.D.L. Rudakov, J.A. Boedo, R.A. Moer et al. Plasma Phys. Control. Fusion 44 (2002) 717 – 731

3. S.I. Lashkul, V.N. Budnikov et al. Plasma Physics Reports.Vol.27 No. 12. 2002 pp.1001-1010

4. S.I. Lashkul et al.. Czechoslovak Journal of Physics (2002) 52 (10) 1149-1159

5. T. Kurki-Suonio et al.. Plasma Phys. and Contr. Fusion 44 (2002) 301

6. V. Rozhansky and M. Tendler, Phys. Fluids B 7 (1992) 1877.

7. T. Kurki-Suonio et al.. 30$^{th}$ EPS Conf. on Contr. Fus. and Pl. Phys. 2003 V.27, P3.145

8. S.V.Shatalin, E.O.Vekshina et al. Plasma Physics Reports. Vol. 30 No. 5. 2004 363-369

9. S.V.Shatalin et al. 30$^{th}$ EPS Conf. on Contr. Fus. and Pl. Phys. 2003 Vol.27, P-3.175

10.S.I. Lashkul et al. 30$^{th}$ EPS Conf. on Contr. Fus. and Pl.Phys. 2003 V.27,P-3.151

11. S.I. Lashkul, V.N. Budnikov at al, Plasma Phys. and Contr. Fus. 44 (2002) 653-663

12. A.B. Altukhov et al., 30$^{th}$ EPS Conf. on Contr. Fus. and Pl.Phys.2003 **27A** , P-4.170pd.

13. Akira Hasegawa and Carol G. Maclennan Phys.Fluids 22(11), November 1979